\documentclass[12pt]{article}
\usepackage{epsfig}

\begin{document}

\vskip 2cm
\title{Domain wall interacting with a black hole: \\
A new example of critical phenomena}
\author{\\
V.P.  Frolov${}^{*} {}^{1}$\,
  A.L. Larsen${}^{|} {}^{2}$ and M. Christensen${}^{||} {}^{2}$}
\maketitle
\noindent
$^{1}${ \em
Theoretical Physics Institute, Department of Physics, \ University of
Alberta, Edmonton, Canada T6G 2J1}
\\$^{2}${\em Physics Department, University of Odense, Campusvej 55, 5230
Odense M,
Denmark}
\vskip 6cm
\noindent
$^{*}$Electronic address: frolov@phys.ualberta.ca\\
$^{|}$Electronic address: all@fysik.ou.dk\\
$^{||}$Electronic address: mc@bose.fys.ou.dk

\newpage
\begin{abstract}
\baselineskip=1.5em
\hspace*{-6mm}
We  study a simple system that comprises all
main features of critical gravitational
collapse,  originally discovered by Choptuik and
discussed in many subsequent publications.
These features include universality of
phenomena, mass-scaling relations,
self-similarity, symmetry between super-critical and sub-critical
solutions, etc.

The system we consider is a  stationary membrane (representing a domain wall)
in a static gravitational field of a black hole. For a membrane that spreads
to infinity, the induced 2+1 dimensional geometry is asymptotically flat.
Besides solutions with Minkowski topology there exists also solutions
with the induced metric and topology of a 2+1 dimensional
black hole. By changing boundary conditions at infinity one finds
that there is a transition between these two families. This transition
is critical and it possesses all the above-mentioned
properties of critical gravitational collapse.
It is remarkable that characteristics of this transition
can be obtained analytically. In particular, we find exact analytical
expressions for scaling exponents and wiggle-periods.

Our results imply that  black hole formation
as a critical  phenomenon is far more general than one might expect.
\end{abstract}

\newpage
\section{Introduction}
\label{intro}
\setcounter{equation}{0}

In our previous publication \cite{bubbles} we considered
stationary axially symmetric membranes (infinitely thin domain walls)
embedded in the
background of a Schwarzschild black hole.
In the approximation when the gravitational field of a membrane
can be neglected, the test membrane configuration is an extremal
of the Dirac-Nambu-Goto action.
Solutions for a membrane which spreads to spatial infinity
form a one parameter family.
We showed that three different
2+1 dimensional membrane
topologies were possible depending on the boundary conditions at infinity:
Minkowski
topology, wormhole topology and
black hole topology.  Moreover, we
found that
the three different membrane topologies are connected via phase transitions
of the form
first discussed by Choptuik \cite{chop}, in investigations of scalar field
collapse. Besides the first order transition (with mass gap)
between black-hole and wormhole topologies, there exists also
 a second order phase transition (no mass gap) between
membranes of
Minkowski topology and membranes of black hole topology. The induced
metric on the intermediate
membrane for the latter transition
has a naked
singularity. For the membranes of black hole topology we found a
mass-scaling relation
analogous to that originally found by Choptuik \cite{chop} (for a review,
see for
instance \cite{gund} and references given therein). We showed that
$Mass\propto
p^\gamma$ where $p$ is an external parameter and $\gamma\approx 0.66$, and
we also found
a periodic wiggle in the scaling relation with period $\omega\approx 3.56$.

The one-parameter family of stationary axially symmetric membranes in the
Schwarzschild
background thus essentially contains all the features of critical
collapse in Einstein theory of gravity,
as have recently been discussed in the
literature (see \cite{gund} for a review). However, there are two important
differences in the
physical setup: (1) In the case of membranes we are
not actually considering the dynamical
formation of
black holes; we are merely considering a one-parameter family of stationary
membranes of
which some have Minkowski topology while others have black hole topology.
(2) In
the case of membranes the induced 2+1 dimensional metric does not obey
Einstein  equations but is determined by solving  the
Dirac-Nambu-Goto equations.

Taking these differences into account, our results \cite{bubbles} indicate
 that
black hole
formation as a critical phenomenon is far more general than originally
expected.

In the present paper, we continue the study of a test stationary
membrane (representing a
thin domain wall) interacting with a generic static black hole and demonstrate
the universality of critical behavior in this system.
The main observation is that all  important features of this critical
behavior are determined only by properties of the membrane solution in the
close vicinity of the event horizon. The latter allows complete
analytical study. We start by considering
 a one-parameter family of uniformly accelerating
axially symmetric membranes in flat Minkowski space,  and demonstrate that
already this very
simple system contains  all the features discussed above.
We use the
results of analytical study of uniformly accelerated membranes to
obtain the characteristics of critical behavior in the general case
of a membrane interacting with a black hole.

The paper is organized as follows. In section \ref{memsec},
we study membranes in Rindler space and describe a
one-parameter family of  membranes which are stationary in a Rindler frame.
We also
discuss some general
properties of these solutions. In
section \ref{critsec}, we
give a different mathematical description of the problem in terms of
phase-portrait and critical
points. This will shed light on some general properties of the membrane
solutions.
In section \ref{analyticsec} we derive analytically
the so-called mass-scaling relation,
that is, the exact
expression of the ``2+1 dimensional mass" of the black hole topology
membranes in terms of
an external ``impact''
parameter. In section \ref{lastsec},
we return to the problem of stationary membranes in the
black-hole background \cite{bubbles}. We demonstrate that
mass-scaling
relation is valid, and obtain  {\it exact} analytical expressions  for the
scaling exponent
and the wiggle-period. Finally in section \ref{conclsec},
we discuss the obtained results.

\section{Stationary Membranes in Rindler Frame}
\label{memsec}
\setcounter{equation}{0}

Our starting point is  the Dirac-Nambu-Goto action
\cite{goto}
\begin{equation}
S \: = \: \mu \int d^3\zeta \sqrt{-\mbox{det}G_{AB}}
\label{action}
\end{equation}
describing a test membrane in an external gravitational field $g_{\mu\nu}$.
Here $\mu$ is the membrane tension, and
\begin{equation}
G_{AB} \: = \: g_{\mu\nu}X^\mu_{,A} X^\nu_{,B}
\label{indmetdef}
\end{equation}
the induced metric on the world-volume.
We denote by $X^\mu$ ($\mu=0,1,2,3$)  spacetime
coordinates, while $\zeta^A\;(A=0,1,2)$ are coordinates on the membrane
world-volume.

We  consider first  axially symmetric membranes in
flat Minkowski space which are
stationary in a Rindler frame, that is, stationary as seen by a uniformly
accelerating observer in Minkowski space. The line-element of Minkowski
space in Rindler frame is
\begin{equation}
ds^2 \: = \: -\alpha ^2Z^2 dT^2+dZ^2+dR^2+R^2d\phi^2 \, ,
\label{rindlermetric}
\end{equation}
where $\alpha^{-1}$ is the acceleration in the $Z$-direction.
The stationary axially symmetric membranes are parametrized as follows
\begin{equation}
T \: = \: \tau  \;\;\; ,\;\;\;
\phi \: =\: \sigma \;\;\; ,\;\;\;
R \: = \: R(Z)\, .
\end{equation}
For such membranes the equations of motion,
corresponding to the
action (\ref{action}),
reduce to the following second order non-linear ordinary
differential
equation
\begin{equation}
ZR R''+(RR'-Z)(1+R'^2) \: = \: 0 \, ,\label{eom}
\end{equation}
where prime denotes derivative with respect to $Z$.
A general solution of this equation  depends on 2 arbitrary constants.
Points where $Z=0$ or $R=0$ are singular points of the equation.
We shall see later that solutions which are regular at these singular points
form a 1-parameter family.

Equation (\ref{eom}) is non-integrable. To demonstrate this, we
introduce parametrization
$Z=x(s),\;R=y(s)$, where
the parameter $s$ is defined by
\begin{equation}
\left( \frac{dx}{ds}\right)^2 +\left( \frac{dy}{ds}\right)^2 \: =
\: x^2 y^2\, .
\label{chaos1}
\end{equation}
Then eqs. (\ref{eom}), (\ref{chaos1}) become
\begin{eqnarray}
\frac{d^2x}{ds^2} &=&y^2x\, ,\nonumber\\
\frac{d^2y}{ds^2}&=&x^2y \, ,\label{chaos2}
\end{eqnarray}
which are exactly the Hamilton equations corresponding to the Hamiltonian
\begin{equation}
H \: = \: \frac{1}{2}\left( p_x^2+p_y^2\right) -\frac{1}{2}x^2 y^2\, .
\label{chaos3}
\end{equation}
A Hamiltonian similar to (\ref{chaos3})
but with the opposite sign of the potential,
is well-known
in the literature in connection with Yang-Mills equations \cite{YM1, YM2,
YM3}, and the
corresponding dynamics has in fact been shown to be chaotic \cite{YM1, YM2,
YM3}.
The system is
thus non-integrable, and the change of sign in the potential does not
affect this
property.

A general solution to eq. (\ref{eom})
can therefore not be obtained in explicit
form, but a
very simple special solution is provided by
\begin{equation}
R \: = \: Z \, .\label{critsol}
\end{equation}
This solution will turn out to play a very special role in our context
of membranes,
2+1 dimensional
black holes and critical phenomena. Moreover, it is easy to show that
the asymptotic
$(Z\rightarrow\infty)$ behavior of a solution of (\ref{eom}) is given by
\begin{equation}
R = Z+\frac{\alpha^{-3/2}}{\sqrt{Z}}\left[
a\cos{ \left( \frac{\sqrt{7}}{2}\ln{(\alpha Z)}\right) } + b\sin{\left(
\frac{\sqrt{7}}{2}\ln{(\alpha Z)}\right) }\right] \, ,\label{asymbeh}
\end{equation}
where $(a,b)$ are dimensionless constants. A membrane is uniquely specified
by its
asymptotic behavior, that is by the 2-vector
\begin{equation}
\vec{p} \: =\: \left( \begin{array}{cc}
a \\ b  \end{array} \right) \, .\label{abdef}
\end{equation}
For solutions which are regular at the singular
points $Z=0$ or $R=0$ of the equation (\ref{eom})
the constants $a$ and $b$ are not independent.

\begin{figure}[htbp]
\caption{Stationary and axially symmetric cosmic membranes in Rindler space
fall into two families: Those of 2+1 dimensional Minkowski spacetime topology
and those of 2+1 dimensional black hole spacetime topology.
The limiting membrane $R=Z$ has a curvature
singularity at $(0,0)$. To obtain the full spatial structure of the membranes,
the curves must be rotated around the $Z$-axis.}
\noindent
\centerline{\psfig{file=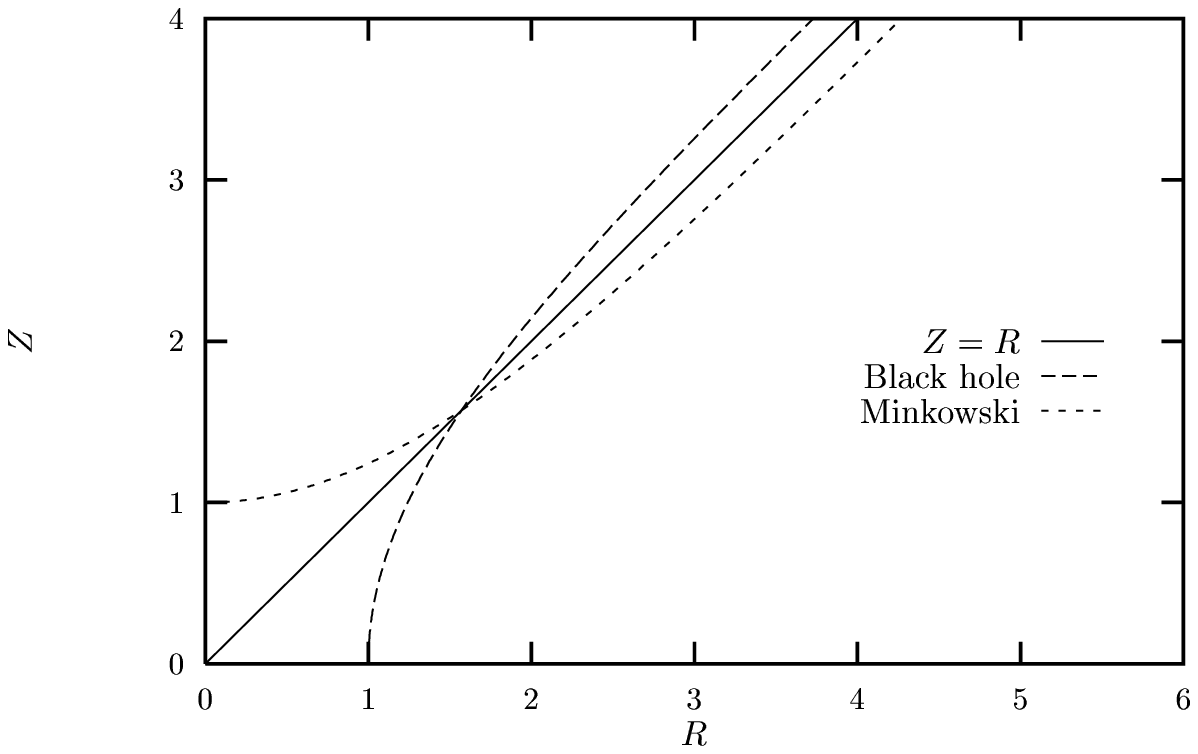,height=30cm,angle=0}}
\label{solfig}
\end{figure}

\begin{figure}[htbp]
\caption{The Minkowski topology and black hole topology solutions oscillate
around the critical solution with a fixed period in a logarithmic plot. Notice
the symmetry relating the two topologies.}
\noindent
\centerline{\psfig{file=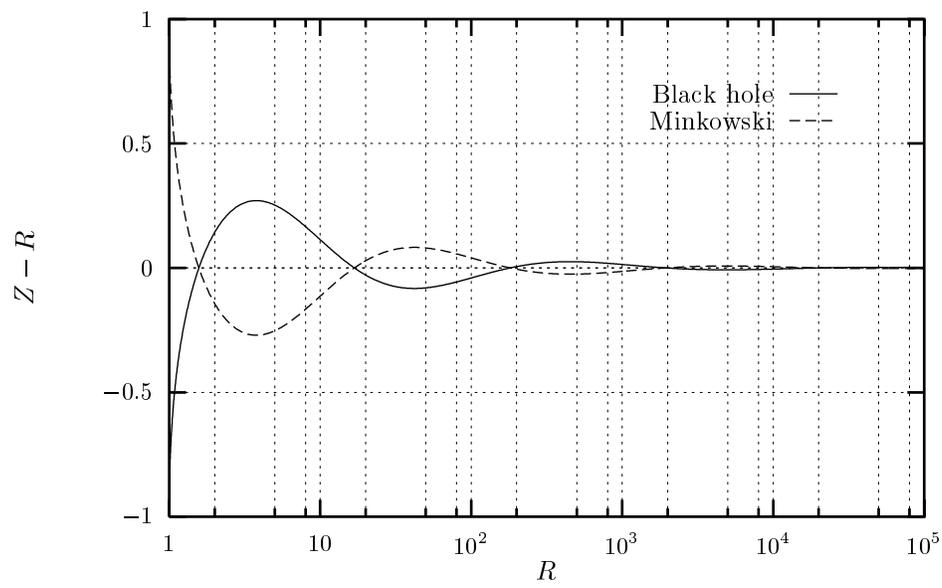,height=30cm,angle=0}}
\label{sollogfig}
\end{figure}

\begin{figure}[htbp]
\caption{Conformal magnification of the parameter space near the critical
solution. This illustrates how
the membranes of the two topologies
approach the critical solution along logarithmic
spiral arms.}
\noindent
\centerline{\psfig{file=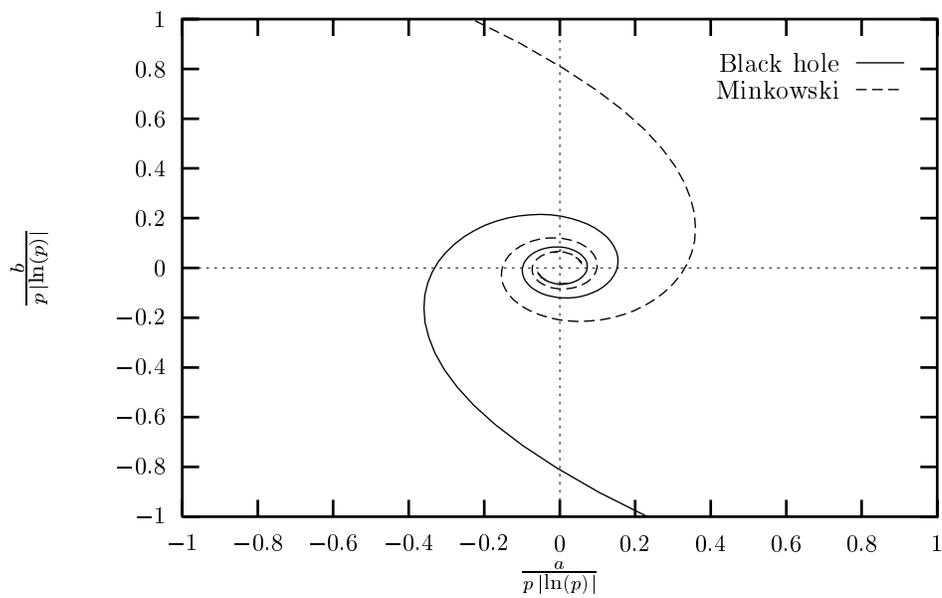,height=30cm,angle=0}}
\label{abfig}
\end{figure}

To demonstrate this,  let us consider the induced metric
on the
membrane world-volume
\begin{equation}
d\Sigma^2 \: = \: -\alpha^2 Z^2d\tau^2+(1+R'^2)dZ^2+R^2d\sigma^2 \, ,
\label{indmet}
\end{equation}
where $R=R(Z)$.  The scalar curvature
corresponding
to the 2+1 dimensional geometries (\ref{indmet})
is
\begin{equation}
{\cal{R}} \: = \:
\frac{-1}{R^2Z^2}\left( \frac{Z^2+R^2R'^2+(RR'-Z)^2}{1+R'^2}\right)\, ,
\end{equation}
so that ${\cal{R}}<0$ but asymptotically ${\cal{R}}\rightarrow 0_-$.
In other words, the induced geometry is asymptotically flat at
infinity $R\rightarrow \infty$.
The induced curvature ${\cal{R}}$ diverges at points where
either $Z=0$ or $R=0$ unless the following conditions are satisfied
\begin{equation}
{\bf{(I)}} \;\;\;\;\;R=0\;\;\; ,\;\;\;\frac{dZ}{dR}=0 \, ;\label{bci}
\end{equation}
\begin{equation}
{\bf{(II)}} \;\;\;\;\;Z=0\;\;\; ,\;\;\;\frac{dR}{dZ}=0 \, .\label{bcii}
\end{equation}
These boundary conditions single out {\em regular}
stationary (for a Rindler observer)
axially symmetric membranes in Minkowski space.
Imposing these conditions reduces the  number of independent parameters
characterizing a solution from 2 to 1.
Some examples of solutions from the 1-parameter family of regular solutions
are shown in
figs. \ref{solfig}, \ref{sollogfig}, and the parameter space is shown in
fig. \ref{abfig}.

Membranes obeying  boundary condition (\ref{bci})
have the topology of 2+1 dimensional Minkowski
spacetime, and can be
thought of as deformed versions of a planar membrane.

An induced geometry for a membrane obeying boundary condition $R(0)=R_{ah}$
for some
$R_{ah}>0$, (\ref{bcii}),   has an apparent horizon at $Z=0$.
These solutions describe a 2+1 dimensional black hole with  ``area''
\begin{equation}
A_{ah} \: = \: 2\pi R_{ah} \, .\label{apphor}
\end{equation}

An exceptional case is
provided by the  particular solution (\ref{critsol}).
This membrane, which is the
limiting solution
separating the Minkowski topology membranes from the black hole topology
membranes, has
the topology of a cone, and its induced metric
has a naked singularity at $(R,Z)=(0,0)$.

Comparing with the results of the critical collapse in
3+1 dimensional Einstein gravity \cite{chop, gund, chop2, brady},
the results of this section already
suggest a close
connection. In particular, the ``critical" membrane solution (\ref{critsol}),
separating the
Minkowski topology membranes from the black hole topology membranes, has a
naked
singularity and it is continuously self-similar in the sense that
\begin{equation}
{\cal L}_{\xi}G_{AB} \: = \: 2G_{AB}\, ,
\label{selfsim}
\end{equation}
where $G_{AB}$ is the induced metric (\ref{indmetdef}) for the
critical solution (\ref{critsol}),
and the vector
$\xi$ is given by
\begin{equation}
\xi^A \: = \: \left( \begin{array}{ccc} 0 \\ Z \\ 0 \end{array} \right)\, .
\label{selfsimvec}
\end{equation}
For the  membranes with black hole topology, it is convenient to use as the
``external parameter"
\begin{equation}
p\equiv |\vec{p}|=\sqrt{a^2+b^2}
\label{paramdef}
\end{equation}
With this definition, the critical solution (\ref{critsol})
corresponds to $p=0$, and
$p$ is a
monotonically decreasing
parameter when approaching the critical solution; see fig. \ref{abfig}.

The goal is now to obtain the so-called ``mass-scaling" relation, i.e. the
relation between the
apparent horizon area (\ref{apphor}) and the parameter $p$ for the black hole
topology membranes. It should
be stressed that, as in the case of scalar field collapse \cite{chop}, the
precise form of the
mass-scaling relation obviously depends on the precise definition of the
``external
parameter" $p$. In the
present case we find that, due to the asymptotic form (\ref{asymbeh}), the
definition (\ref{paramdef}) is the most
natural. In the case of stationary membranes in the Schwarzschild
background \cite{bubbles}, a similar
definition was used.

\section{Phase-Portrait and Critical Points}
\label{critsec}
\setcounter{equation}{0}

Before deriving mass-scaling relation,
we give an alternative mathematical description of the
differential equation (\ref{eom}).

\begin{figure}[htbp]
\caption{Phase-portrait of the system of equations (\ref{auto}). It has a
saddle point at $(0,1/2)$ and a focus point at $(1,1)$. The curve connecting
these two points represents all the black hole topology membranes. }
\noindent
\centerline{\psfig{file=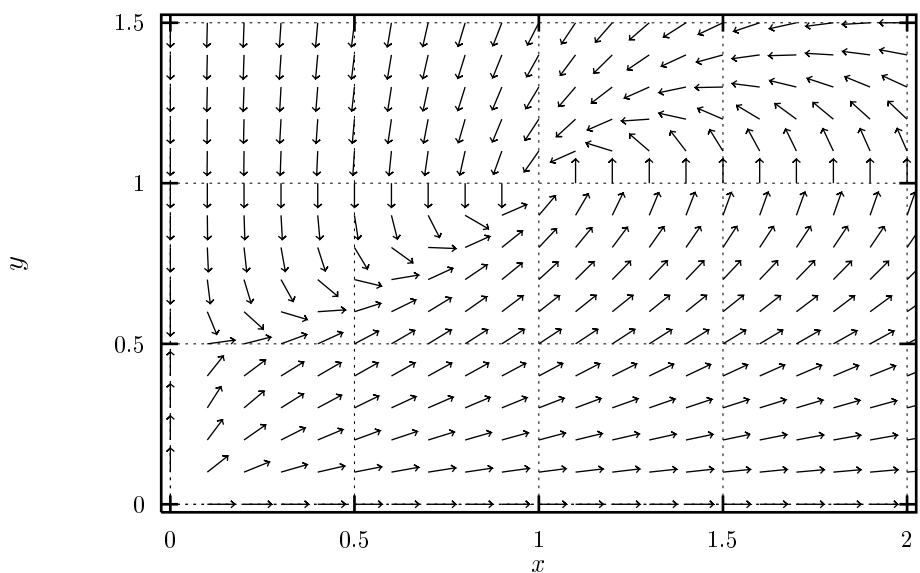,height=30cm,angle=0}}
\label{phasefig}
\end{figure}

It is convenient to introduce new coordinates $(x,y)$ (not to be confused with
those of eqs. (\ref{chaos1})-(\ref{chaos3})) by
\begin{equation}
x  =  R' \, ,\hspace{1cm}
y  =  \frac{R}{Z} \, R' \, ,\label{newxy}
\end{equation}
as well as a new parameter $s$
\begin{equation}
\frac{dZ}{Z} \: = \: y(s) \, ds \, .
\end{equation}
Then eq. (\ref{eom}) becomes a first order regular autonomous system
\begin{eqnarray}
\frac{dx}{ds} & = & x \, (1-y) \, (1+x^2) \, ,\nonumber \\
\frac{dy}{ds} & = & y \, \left( 1 - 2y + x^2 \, (2-y) \right)\, .
\label{auto}
\end{eqnarray}
The corresponding phase-portrait is shown in fig. \ref{phasefig}.
Notice that there are four critical points;
\begin{eqnarray}
\left( \begin{array}{cc} x \\ y \end{array} \right) & = &
\left( \begin{array}{cc} 0 \\ 0 \end{array} \right)
\; \; \; \; \; \; \; \; \; \:
\mbox{node} \label{unnode} \\
\left( \begin{array}{cc} x \\ y \end{array} \right) & = &
\left( \begin{array}{cc} 0 \\ \frac{1}{2} \end{array} \right)
\; \; \; \; \; \; \; \; \; \:
\mbox{saddle point} \label{saddle} \\
\left( \begin{array}{cc} x \\ y \end{array} \right) & = &
\left( \begin{array}{cc} \pm 1 \\ 1 \end{array} \right)
\; \; \; \; \; \; \;
\mbox{focus points} \label{focus}
\end{eqnarray}
one being a node, another being a saddle point
and the last two are focus points.

The focus points (\ref{focus}), which in the phase-portrait are approached
by self-similar
(logarithmic) spirals, correspond to the two critical membranes $R=\pm Z$.
Constraining ourselves to the one half part of spacetime $Z>0$ lying above the
Rindler horizon, only the focus with positive $x$-component is relevant.

The saddle point (\ref{saddle}) precisely corresponds to the
boundary condition
(\ref{bcii}) for the membranes of 2+1 dimensional black hole topology, that is
\begin{equation}
R'=0 \; \; \; , \; \; \; Z=0 \; \; \; , \; \; \; R \; \mbox{arbitrary}\, .
\end{equation}
The important consequence is that, when using these $(x,y)$-coordinates, {\it
all} membranes of 2+1 dimensional
black hole topology are represented by only one curve,
namely the curve connecting the saddle point $(0,1/2)$ and the focus point
$(1,1)$. So it is
necessary to make only one numerical integration of the system (\ref{auto});
then {\it all}
membranes of 2+1 dimensional black hole topology
can be reconstructed. In the next
section we shall see that this fact
is closely related to a certain symmetry of the equation (\ref{eom}).

Notice also that the  node $(0,0)$ is
represented neither by the critical solution nor by any other
solution obtained from the boundary conditions (\ref{bci}), (\ref{bcii}) and
hence it has no relevance here.

\section{Analytical Approach to Mass-Scaling}
\label{analyticsec}
\setcounter{equation}{0}

Now return to the original equation (\ref{eom}). There are two symmetries that
will allow us to give a complete analytical description of the critical
phenomena concerning the transition between membranes of
2+1 dimensional Minkowski
topology and membranes of 2+1 dimensional black hole topology.

Notice first that the boundary conditions (\ref{bci}), (\ref{bcii}) are
symmetric under interchange of $R$ and $Z$
\begin{equation}
R\: \leftrightarrow \: Z\, .\label{sym1}
\end{equation}
This symmetry actually extends to
the solutions themselves as follows from eq. (\ref{eom}) (see also
eq. (\ref{chaos2})). In our context it means that for any Minkowski topology
membrane $R=F(Z)$, there is a corresponding black hole topology membrane
$Z=F(R)$ with the same function $F$ (see Figure 1).

It is also easy to see that if $R(Z)$ is a solution to eq. (\ref{eom})
then, for arbitrary $k >0$, $R(kZ)/k$ is also a solution
\begin{equation}
R(Z) \: \leftrightarrow \: \frac{R(kZ)}{k}\, . \label{sym2}
\end{equation}
Transformation (\ref{sym2}) preserves  boundary condition (\ref{bcii}),
but it shifts the apparent horizon (\ref{apphor}). Thus, from one black hole
topology membrane, we can construct all the others by the transformation
(\ref{sym2}). Using the transformation (\ref{sym1}), we can then also construct
all the Minkowski topology membranes corresponding to the boundary conditions
(\ref{bci}).

To make these statements more precise, consider some fixed
``reference-solution"
$\tilde R (Z)$, corresponding to a black hole topology membrane with apparent
horizon $\tilde R _{ah}$ and asymptotic behavior (\ref{asymbeh}) with constants
$(\tilde a , \tilde b )$, that is
\begin{equation}
\tilde R (0) \: = \: \tilde R _{ah}\, , \; \; \; \; \;  \; \; \; \; \; \;
\tilde R '(0) \: = \: 0\, ,
\end{equation}
and for $Z \rightarrow \infty$
\begin{equation}
\tilde R (Z) = Z+\frac{\alpha^{-3/2}}{\sqrt{Z}}\left[
\tilde a \cos{ \left( \frac{\sqrt{7}}{2}\ln{(\alpha Z)}\right) } +
\tilde b \sin{\left(
\frac{\sqrt{7}}{2}\ln{(\alpha Z)}\right) }\right] \, .
\end{equation}
Any other black-hole-topology membrane $R(Z)$ can then be generated by the
transformation (\ref{sym2}) for some $k>0$. This new solution is
characterized by
\begin{equation}
R (0) \: = \: \frac{\tilde R (0)}{k} \: = \: \frac{\tilde R _{ah}}{k} \:
\equiv \: R_{ah} \, ,\; \; \; \; \;  \; \; \; \; \; \; R '(0) \: = \: 0\, ,
\label{transbc}
\end{equation}
and for $Z \rightarrow \infty$
\begin{eqnarray}
R (Z) & = & Z+\frac{\alpha^{-3/2}}{k^{3/2} \sqrt{Z}}\left[
\tilde a \cos{ \left( \frac{\sqrt{7}}{2}\ln{(\alpha kZ)}\right) } +
\tilde b \sin{\left(
\frac{\sqrt{7}}{2}\ln{(\alpha kZ)}\right) }\right] \nonumber \\
& \equiv & Z+\frac{\alpha^{-3/2}}{\sqrt{Z}}\left[
a \cos{ \left( \frac{\sqrt{7}}{2}\ln{(\alpha Z)}\right) } +
b \sin{\left(
\frac{\sqrt{7}}{2}\ln{(\alpha Z)}\right) }\right]\, .
\end{eqnarray}
Here the second line defines the constants $(a,b)$ according to eq.
(\ref{asymbeh}).
It follows that
\begin{eqnarray}
a & = & k^{-3/2} \left[
\tilde a \cos{ \left( \frac{\sqrt{7}}{2}\ln{(k)}\right) }
+ \tilde b \sin{\left( \frac{\sqrt{7}}{2}\ln{(k)}\right) }
\right] \, ,\nonumber \\
b & = & k^{-3/2} \left[
\tilde b \cos{ \left( \frac{\sqrt{7}}{2}\ln{(k)}\right) }
- \tilde a \sin{\left( \frac{\sqrt{7}}{2}\ln{(k)}\right) }
\right] \, .\label{transpiral}
\end{eqnarray}
This shows that we can generate the complete parameter space, fig. \ref{abfig},
from the ``reference point" $(\tilde a,\tilde b)$, and that the parameter space
consists of two logarithmic spirals. Notice that the points corresponding to
Minkowski topology membranes are obtained from the points corresponding to
black hole topology membranes by inversion $(a,b) \leftrightarrow (-a,-b)$, as
follows from eq. (\ref{sym1}).

However, we can go one step further. Consider again the relations
(\ref{transbc}), (\ref{transpiral}) for the black hole topology membranes.
The transformation
(\ref{transpiral}) is a combined scaling and rotation in parameter space. Since
\begin{equation}
k=\frac{{\tilde R}_{ah}}{R_{ah}}=\frac{{\tilde A}_{ah}}{A_{ah}}\, ,
\end{equation}
the rotation matrix can be written as a product of two rotation matrices;
one involving only
$\tilde{A}_{ah}$ and the other involving only $A_{ah}$. That is,
relations (\ref{transpiral})
are equivalent to
\begin{eqnarray}
&A_{ah}^{-3/2}&\left( \begin{array}{cc} \cos(\frac{\sqrt{7}}{2}\ln{(A_{ah})}) &
 \sin(\frac{\sqrt{7}}{2}\ln{(A_{ah})}) \\
-\sin(\frac{\sqrt{7}}{2}\ln{(A_{ah})}) &
\cos(\frac{\sqrt{7}}{2}\ln{(A_{ah})}) \, ,\end{array}
\right) \left( \begin{array}{cc} a \\ b \end{array}
\right) \nonumber\\ &=& \tilde{A}_{ah}^{-3/2}\left( \begin{array}{cc}
\cos(\frac{\sqrt{7}}{2}\ln{(\tilde{A}_{ah})}) &
 \sin(\frac{\sqrt{7}}{2}\ln{(\tilde{A}_{ah})}) \\
-\sin(\frac{\sqrt{7}}{2}\ln{(\tilde{A}_{ah})}) &
\cos(\frac{\sqrt{7}}{2}\ln{(\tilde{A}_{ah})}) \end{array}
\right) \left( \begin{array}{cc} \tilde{a} \\ \tilde{b} \end{array}
\right) \, ,\label{rotsplit}
\end{eqnarray}
where the horizon area is measured in units of $\alpha^{-1}$. It follows
that the combination:
\begin{eqnarray}
A_{ah}^{-3/2}\left( \begin{array}{cc} \cos(\frac{\sqrt{7}}{2}\ln{(A_{ah})}) &
 \sin(\frac{\sqrt{7}}{2}\ln{(A_{ah})}) \\
-\sin(\frac{\sqrt{7}}{2}\ln{(A_{ah})}) &
\cos(\frac{\sqrt{7}}{2}\ln{(A_{ah})}) \end{array}
\right) \left( \begin{array}{cc} a \\ b \end{array}
\right) \label{rindlerconstvec}
\end{eqnarray}
is a constant vector, i.e. the same for {\it all} membranes of black hole
topology. Taking the norm and
using definition (\ref{paramdef}) finally leads to:
\begin{equation}
\frac{p^2}{A_{ah}^3} \: = \: {\mbox{const}}\, .
\label{constfrac}
\end{equation}
It follows that we get the exact mass-scaling relation
\begin{equation}
\ln \left( A_{ah} \right) \: = \:
\frac{2}{3} \, \ln \left( p \right) + c \label{massscale}
\label{rindlermassscale}
\end{equation}
corresponding to a critical exponent $\gamma=2/3$, and $c$ is an
unimportant constant.
This exactly linear mass-scaling relation without any wiggle was
to be expected due to the continuous self-similarity
(\ref{selfsim}), (\ref{selfsimvec}), if we believe in the analogy with the
case of scalar field
collapse \cite{chop, gund2, hod}. The relation (\ref{constfrac})
also implies that the ``phase-transition"
between membranes of Minkowski topology and membranes of black hole
topology is of second order (no mass gap).

In this section we have obtained the mass-scaling relation for the black hole
topology membranes,
that is, the super-critical solutions. However, since we have the exact
symmetry (\ref{sym1}), we
could as well have obtained the scaling relation for the Minkowski topology
membranes, that is, the sub-critical solutions. In that case one would use
instead of the
apparent horizon area (\ref{apphor}), the minimal
proper distance between the Minkowski
topology membrane and the Rindler horizon. The scaling relations in the two
cases would then
obviously be identical. It is interesting to mention that such "symmetry"
between sub-critical
and super-critical scaling relations have been found numerically in the
case of scalar field
collapse also \cite{garfinkle}. In our case, this symmetry is completely
trivial.

\section{Black Hole Background Case}
\label{lastsec}
\setcounter{equation}{0}

In a previous publication \cite{bubbles}, we considered stationary axially
symmetric membranes in the Schwarzschild spacetime
\begin{equation}
ds^2 \: = \: - \left( 1-\frac{2M}{r} \right) dt^2
+ \left( 1-\frac{2M}{r} \right) ^{-1} dr^2 + r^2 \left( d \theta ^2 +
\sin ^2 \theta d \phi ^2 \right)\, .
\label{schmetric}
\end{equation}
Such membranes are parametrized by
\begin{equation}
t \: = \: \tau \, ,
\; \; \;  \; \; \; \phi \: = \: \sigma \, ,
\; \; \;  \; \; \; \theta \: = \: \theta (r)\, ,
\end{equation}
and the equation of motion corresponding to the action (\ref{action}) is
\begin{equation}
\theta '' + (2r - 3M) {\theta '}^3 - \frac{{\theta '}^2}{\tan (\theta )}
+ \frac{3r-4M}{r(r-2M)} \, \theta ' - \frac{1}{r(r-2M) \, \tan (\theta )} \: =
\: 0\, .
\label{schembed}
\end{equation}
where the prime now denotes derivative with respect to $r$.

In reference \cite{bubbles} we showed that three different membrane topologies
were possible depending on the boundary conditions at infinity:
2+1 dimensional Minkowski
topology, 2+1 dimensional wormhole topology
and 2+1 dimensional black hole topology.
We also found that the different membrane topologies are connected via phase
transitions of the form first discussed by Choptuik
\cite{chop} in investigations
of scalar field collapse. In particular, we found a second order phase
transition (no mass gap) between membranes of Minkowski topology and
membranes of black hole
topology. The corresponding mass-scaling relation for the black hole topology
membranes was numerically found to have the approximate form
\begin{equation}
\ln(Mass) \: \approx \: \gamma \ln (p) + f(\ln (p))\, ,
\end{equation}
where $p$ is the external parameter and $f$ is a periodic function with
period $\omega $.
The numerical computations gave \cite{bubbles}
\begin{equation}
\gamma  \approx  0.66 \, ,\hspace{1cm}
\omega  \approx  3.56 \, .
\label{formerres}
\end{equation}
We shall now show analytically that $\gamma = 2/3$ and $\omega = 3 \pi
/\sqrt{7}$, and that these numbers are universal, in the sense that they hold
true for stationary axially symmetric membranes in a variety of background
spacetimes, including Schwarzschild, Reissner-Nordstr\"om and Rindler
spacetimes (in the latter case, the periodic function has zero amplitude, as
will be discussed later). \\

Our main observation is that characteristics of critical behavior are
determined only by properties of a membrane in the close vicinity of
the event horizon of the background geometry. To illustrate this,
we consider black-hole-topology membranes in the Schwarzschild
background, although the arguments are more general.
It is well-known that the
near-horizon region of a Schwarzschild (as of any other static)
black hole is Rindler space. To be more
precise, consider the region  near the $\theta=\pi$-pole: $r=2M+\xi$ ,
$\theta = \pi -\eta$,
of the Schwarzschild black hole, and define
\begin{equation}
Z \: = \: \sqrt{8 M \xi}\, , \; \; \;  \; \; \; R \: = \: 2 M \eta \, .
\end{equation}
To the lowest order the line element (\ref{schmetric}) then becomes
\begin{equation}
ds^2 \: = \: - \, \frac{Z^2}{16 M^2} dt^2 + dZ^2 + dR^2 + R^2 d\phi ^2
\, ,\end{equation}
which is precisely the Rindler line element (\ref{rindlermetric})
for $\alpha =
\frac{1}{4M}$. Moreover, the membrane-embedding equation (\ref{schembed})
becomes:
\begin{eqnarray}
ZRR'' + \left( RR' - Z \right) \, \left( 1 + {R'}^2 \right)  &=&
-\frac{Z^2RR'^3}{4M^2} -
\frac{Z^2RR'}{4M^2}\;\frac{1}{1 + (\frac{Z}{4M}) ^2 }   \nonumber\\
 - Z \left( 1 - \frac{R}{2M \tan (\frac{R}{2M})} \,
\frac{1}{1 + (\frac{Z}{4M}) ^2} \right)
&-& Z {R'}^2 \left( 1 - \frac{R}{2M \tan (\frac{R}{2M})}\right)\, ,
\label{schequ}
\end{eqnarray}
which is precisely eq. (\ref{eom}) near the horizon, since the terms on the
right hand side are negligible there.

As in Rindler space, a membrane in the Schwarzschild spacetime is specified by
a 2-vector $\vec p = (a,b)$ which parametrizes the solutions of the embedding
equation at spatial infinity. Near the critical membrane (which separates the
Minkowski topology membranes from the black hole topology membranes), it is
convenient to choose the parametrization such that the critical membrane
corresponds to $(a,b)=(0,0)$ and such that $p \equiv \sqrt{a^2 + b^2}$ is a
decreasing function when approaching the critical solution (see ref.
\cite{bubbles} for more details). In the Schwarzschild background, the
critical membrane enters the horizon at $r=2M$, $\theta = \pi$ where it has a
naked singularity. Near-critical membranes are thus described by equation
(\ref{schequ}), with vanishing right hand side near the horizon.

Now consider such a near-critical membrane of black hole topology; we think of
this solution as being obtained by linearization around the critical membrane.
The idea is then to first read off Rindler-data $(a_R,b_R)$ in the Rindler
region, and then to relate this to the Schwarzschild-data $(a,b)$ which we read
off at spatial infinity. This makes sense since a membrane sufficiently close
to the critical membrane will make a sufficient number of oscillations (see
fig. \ref{sollogfig}) in the Rindler region (this statement actually
follows from the
symmetry (\ref{sym2})) to allow determination
of the coefficients $(a_R,b_R)$, as
defined by eq.
(\ref{asymbeh}). Moreover, since this near-critical membrane is obtained by
linearization
around the critical membrane, there will to lowest order be a linear
relationship between the
Rindler-data and the Schwarzschild-data:
\begin{equation}
\left( \begin{array}{c} a \\ b \end{array} \right) \: = \: {\cal A}
\left( \begin{array}{c} a_R \\ b_R \end{array} \right)
\label{lintrans}
\end{equation}
for some constant 2$\times$2 matrix ${\cal A}$, which somehow have encoded the
spacetime curvature (for ``pure" Rindler space, the matrix ${\cal A}$ is just
the identity). By repeating the argument of section \ref{analyticsec}, it is
now possible to obtain the exact mass-scaling relation for black hole topology
membranes in the Schwarzschild background.

\begin{figure}[htbp]
\caption{A simple fit of the non-inverted mass-scaling relation
(\ref{fitscale})
for the
black hole topology membranes in the Schwarzschild background indicates that
the constant $C_2$ is indeed quite large. Here the values
$C_1 \approx 0.0188$,
$C_2=0.858$ and
$\varphi=4.69$ have been used.  }
\noindent
\centerline{\psfig{file=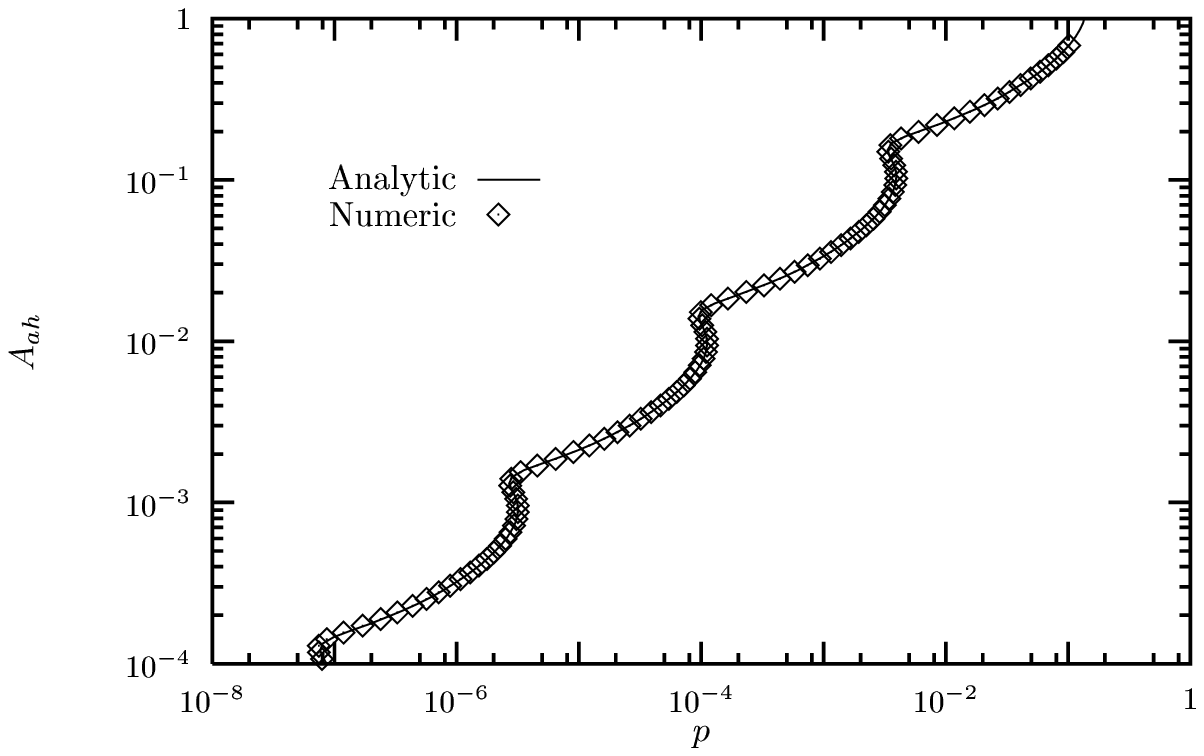,height=30cm,angle=0}}
\label{fitfig}
\end{figure}

Consider again some fixed
``reference-solution" corresponding to a black hole topology membrane with
apparent horizon $\tilde R_{ah}$, as well as another solution corresponding to
the apparent horizon $R_{ah}$. Both solutions are very close to the critical
membrane solution and thus have respective Rindler-data
$(\tilde a_R,\tilde b_R)$ and $(a_R,b_R)$, as well as Schwarzschild-data
$(\tilde a,\tilde b)$ and $(a,b)$. Now using relations (\ref{lintrans}) and
(\ref{transpiral}) we get
\begin{equation}
\left( \begin{array}{c} a \\ b \end{array} \right) \: = \:
k^{-3/2} \! {\cal A}
\left( \begin{array}{cc}
\cos \left( \frac{\sqrt{7}}{2} \ln \left(
k \right) \right) &
\sin \left( \frac{\sqrt{7}}{2} \ln \left(
k \right) \right) \\
- \sin \left( \frac{\sqrt{7}}{2} \ln \left(
k \right) \right) &
\cos \left( \frac{\sqrt{7}}{2} \ln \left(
k \right) \right)
\end{array} \right)
\! {\cal A}^{-1} \!
\left( \begin{array}{c} \tilde a \\ \tilde b \end{array} \right)\, ,
\end{equation}
where $k=\tilde A_{ah}/A_{ah}$. That is, by repeating the analysis of eqs.
(\ref{rotsplit})-(\ref{rindlerconstvec}), a
vector being constant for all near-critical membranes of black hole
topology exists
\begin{equation}
A_{ah}^{-3/2} \, \left( \begin{array}{cc}
\cos \left( \frac{\sqrt{7}}{2} \ln (A_{ah}) \right) &
\sin \left( \frac{\sqrt{7}}{2} \ln (A_{ah}) \right) \\
- \sin \left( \frac{\sqrt{7}}{2} \ln (A_{ah}) \right) &
\cos \left( \frac{\sqrt{7}}{2} \ln (A_{ah}) \right) \end{array} \right)
\, {\cal A}^{-1} \,
\left( \begin{array}{c}  a \\  b \end{array} \right)\, .
\label{schconstvec}
\end{equation}
Taking the norm leads to
\begin{equation}
p^2 \: \equiv \: a^2 + b^2 \: = \: A_{ah}^3 \; C_1 \left[
1 + C_2 \cos \left( \sqrt{7} \, \ln{(A_{ah})} - \varphi \right) \right]\, ,
\end{equation}
where $(C_1,C_2,\varphi )$ are constants depending only on the background
spacetime!
Furthermore, from eq. (\ref{schconstvec}) follows directly
\begin{eqnarray}
\ln (A_{ah}) & = & \frac{2}{3} \ln (p) - \frac{1}{3} \ln (C_1) \nonumber \\
& & - \frac{1}{3} \ln \left[ 1 + C_2
\cos \left( \sqrt{7} \, \ln (A_{ah}) - \varphi \right) \right]\, .
\label{fitscale}
\end{eqnarray}
Now let us compare with the numerical results obtained in ref.
\cite{bubbles}. The
numerically obtained mass-scaling relation is shown in fig. \ref{fitfig},
together with a fit for the parameters $(C_1,C_2,\varphi )$
obtained using eq. (\ref{fitscale}). It follows that there is
indeed complete agreement.

It is sometimes  convenient to write eq. (\ref{fitscale}) in a
different way. The
expression can not be
inverted analytically, but
 since we are near the
critical solution, the first term on the right hand side is dominant (this
is a reasonably
good approximation provided $C_2$ is not too close to $1$). Therefore
\begin{eqnarray}
\ln (A_{ah}) & \approx & \frac{2}{3} \ln (p) - \frac{1}{3} \ln (C_1)
\nonumber \\
& & - \frac{1}{3} \ln \left[ 1 + C_2
\cos \left( \frac{2 \sqrt{7}}{3} \ln (p) - \varphi \right) \right]\, .
\label{r1}
\end{eqnarray}
This relation is of the standard form \cite{gund}
\begin{equation}
\ln (Mass) \: \approx \: \gamma \ln (p) + C + f(\ln (p))\, ,
\label{r2}
\end{equation}
where $f$ is a periodic function; $f(z) = f(z+\omega )$. It must be
stressed, however, that
eq. (\ref{fitscale}) gives a more precise
expression for the mass-scaling relation.
But following eqs.
(\ref{r1})-(\ref{r2}), we find  in our case
\begin{equation}
\gamma  = \frac{2}{3} \, ,\hspace{1cm}
\omega =  \frac{3 \pi}{\sqrt{7}}\, ,
\label{univres}
\end{equation}
in agreement with eq. (\ref{formerres}), while
\begin{eqnarray}
C & = & - \frac{1}{3} \ln (C_1) \, ,\nonumber \\
f(z) & = & - \frac{1}{3} \ln \left[ 1 + C_2 \cos
\left( \frac{2 \sqrt{7}}{3} z - \varphi \right) \right]\, .
\label{wiggleform}
\end{eqnarray}
The constants $(C_1,C_2,\varphi )$ determining the constant $C$ and the
function $f(z)$ can be obtained numerically (c.f. fig. \ref{fitfig}),
but they depend
on the embedding spacetime. The
important point is that the constants $\gamma $ and $\omega $ are
\underline{universal}, that is they have the same values (\ref{univres})
for an arbitrary black-hole background geometry.
 In this respect, membranes in ``pure" Rindler space as
discussed in Sections
\ref{intro}-\ref{analyticsec}, represents a ``degenerate" case. Indeed, in
Rindler space the matrix ${\cal A}$, as introduced in eq. (\ref{lintrans}),
is just the identity matrix,
and therefore the constant $C_2$ in eq. (\ref{wiggleform}) equals 0. Thus
in that case, there
is no periodic wiggle in the scaling relation (more precisely, its amplitude
vanishes) and the result reduces to eq. (\ref{rindlermassscale}).

\section{Conclusion}
\label{conclsec}
\setcounter{equation}{0}

In this paper we have considered stationary cosmic membranes in Rindler
space as well as in
black hole spacetimes, thus generalizing and completing the previously
obtained  results
\cite{bubbles}. Using exact analytical methods, we have analyzed in detail
the transition
between the different membrane topologies. We have shown that this simple
system comprises
{\it all} features of gravitational collapse of scalar and Yang Mills
fields, as originally
discovered and discussed by Choptuik \cite{chop}. These features include
mass-scaling relations (with or without wiggles), universality of phenomena
and certain dimensionless quantities (critical exponent and wiggle-period),
mass gap (or no
mass gap), self-similarity (continuous or discrete), symmetry between
super-critical and
sub-critical solutions, etc.

Using our simple system, we have thus been able
to account for all  these
features in a completely analytical way.
The obtained results are  in complete
agreement with previous
numerical results and explain them.
For instance, we have obtained exact analytical
expressions for scaling
exponents and wiggle-periods.

Our results indicate that  black hole formation
as a critical  phenomenon might be far more general
than expected. In particular,
the critical phenomena of black hole formation are not
restricted to time-dependent solutions of any
particular physical equation, like the Einstein equation.

\vskip 24pt
\hspace*{-6mm}{\bf Acknowledgements}:\\
We would like to thank A.V. Frolov for helpful discussions on critical
phenomena. One of the
authors (V.F.) is grateful to the
Natural Sciences and Engineering Research Council of Canada and
the Killam Trust for their financial
support.

\newpage

\newpage

\newpage

\newpage

\newpage
\begin{thebibliography}{12}

\bibitem{bubbles}M. Christensen, V.P. Frolov and A.L. Larsen, Phys. Rev. D
{\bf 58}, 085005 (1998).

\bibitem{chop}M.W. Choptuik, Phys. Rev. Lett. {\bf 70}, 9 (1993).

\bibitem{gund}C. Gundlach, Adv. Theor. Math. Phys. {\bf 2}, 1 (1998).

\bibitem{goto}P.A.M. Dirac, Proc. R. Soc. London A {\bf 268}, 57 (1962).

\bibitem{YM1}S.G. Matinyan, G.K. Savvidy and N.G. Ter-Arutyunyan-Savvidy,
Sov. Phys. JETP {\bf 53}, 421 (1981).

\bibitem{YM2}G.K. Savvidy, Phys. Lett. B {\bf 130}, 303 (1983) .

\bibitem{YM3}P. Dahlquist and G. Russberg, Phys. Rev. Lett. {\bf 65},
2837 (1990).


\bibitem{chop2}M.W. Choptuik, T. Chmaj and P. Bizon, Phys. Rev. Lett. {\bf
77}, 424 (1996).

\bibitem{brady}P.R. Brady, C.M. Chambers and S.M.C.V. Gon\c{c}alves,
Phys. Rev. D {\bf 56}, R6057 (1997).

\bibitem{gund2}C. Gundlach, Phys. Rev. D {\bf 55}, 695 (1997).

\bibitem{hod}S. Hod and T. Piran, Phys. Rev. D {\bf 55}, 440 (1997).

\bibitem{garfinkle}D. Garfinkle and G.C. Duncan, Phys. Rev. D {\bf 58}, 064024
(1998).

\end{thebibliography}
\end{document}